# The simplest complexity:
# The story of the three-body problem

Barak Kol[1]

The Hebrew University of Jerusalem

**Abstract**

This article offers a broad-brush account of the Newtonian three-body problem, from its origins with Newton to its vibrant present, emphasizing its enduring influence on theoretical physics. It unfolds through a series of self-contained episodes that illuminate the scientific fields and the paradigm shift that have grown out of this problem.

## Table of Contents

[1] barak.kol@mail.huji.ac.il

## Invitation

The three-body problem is one of the richest, deepest and longest-standing open problems in physics. It is the fertile soil that nurtured the paradigm shift from the clockwork universe to chaos, and it grew numerous scientific theories including perturbation theory, the symplectic formulation of mechanics, and the mathematical field of topology.

The problem is easy to state. Consider three point-like bodies moving under the influence of their mutual Newtonian gravitational forces. Given their initial positions and velocities, predict their future motion (Figure 1).

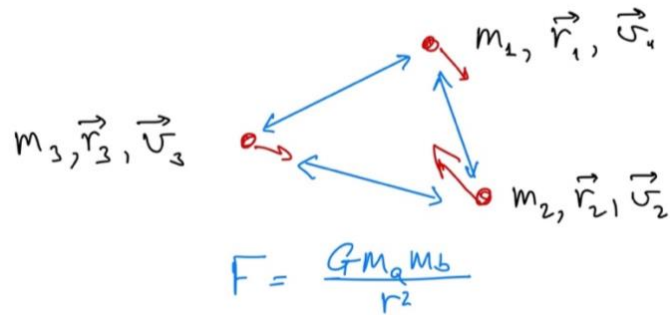

*Figure 1 Definition of the three-body system.*

While the motion of a two-body system can be predicted into the far future and is simple in this sense, the three-body system, its most immediate generalization, displays complex motion and is not easy to solve.

Excellent three-body reviews can be found, including (Gutzwiller, 1998) and (Valtonen & Karttunen, 2006). In this essay, we follow the fascinating scientific story of the three-body problem in the light of recent developments regarding a statistical solution. Along the way, we trace several colorful threads—some concerning the formulation, others its attempted solution, and still others the mathematical ideas it inspired. We shall see in what sense it is one of the simplest systems that defy deterministic prediction, and thereby displays the essence of complexity.

## Search for deterministic prediction

### Newton and the Moon
Our story begins with Isaac Newton. In the Principia (Newton, 1687), he famously explained the motion of the planets around the Sun using his three laws of motion and his law of universal gravitation (Figure 2).

Like every great scientific achievement, it provoked new questions that had not even been conceivable before. Moreover, the fame of the achievement inspired later researchers to apply similar reasoning to a multitude of other problems—inevitably failing for some—thereby sowing the seeds of the following powerful ideas.

In the case of the Principia, the new question made possible was to explain the motion of the Moon, a study that Newton himself initiated. In the crudest approximation, the Moon orbits the Earth in a Keplerian orbit, similar to those of the planets around the Sun. However, the ancients already knew that this orbit displays certain slow drifts. More precisely, the orbital plane precesses (slowly rotates) around an axis perpendicular to the plane of the Solar System (the ecliptic)—a motion known as the nodal precession (Figure 3). In addition, within the orbital plane, the Keplerian ellipse precesses, such that the perigee and apogee, the orbit's closest and farthest points from the Earth, precess—a motion known as apsidal precession. These precessions and their rates appear already in the Almagest (Ptolemy, 150 AD).

Whereas the ancients believed that the Moon and all celestial bodies move on some celestial tracks, Newton's theory stipulates that nothing but the gravitational forces hold the Moon in its orbit. Newton realized that the precession of the Moon's orbit is due to the Sun, which in addition to keeping the Earth–Moon system bound to it, exerts a residual force on the Moon's orbit. In order to explain the motion of the Moon, Newton introduced the three-body problem, in the specific hierarchical case of the Earth–Moon–Sun system; see (Gutzwiller, 1998) for an excellent review of this system. In this context, Newton studied perturbations to Keplerian orbits and was able to explain the nodal precession and part of the apsidal precession (Figure 4). Yet beneath this apparent regularity, the seeds of unpredictability were already sown.

Interestingly, the lunar-like three-body problem occupies a significant part of the Principia (about one eighth of it), while its most famous achievement, the explanation of planetary orbits, is limited to about a third of the book. If we take the view that the Principia is the first work in Physics, we should recognize the study of the lunar precessions to be the first appearance of perturbation theory in physics. In his commentary to the Principia, the remarkable Chandrasekhar finds "one may in truth say that there is hardly anything in any modern textbook on celestial mechanics … that one cannot find in the [Principia], and indeed with deeper understanding" (Chandrasekhar, 1995, p. 444).

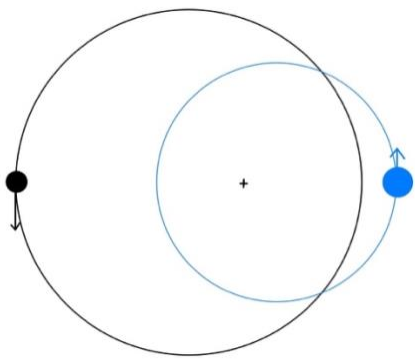
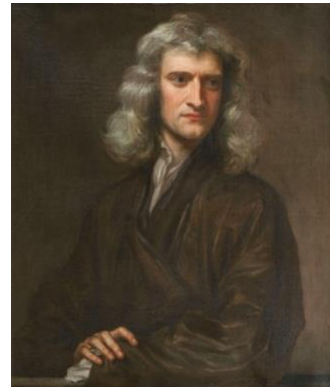

*Figure 2 Newton obtained the exact solution to the gravitational two-body problem: two bodies orbit their common center of mass on ellipses of equal eccentricity. Here, two bodies with a mass ratio of 0.7 are shown moving on elliptical orbits with eccentricity e=0.3. Left panel: Isaac Newton (1642–1727) in 1689.*

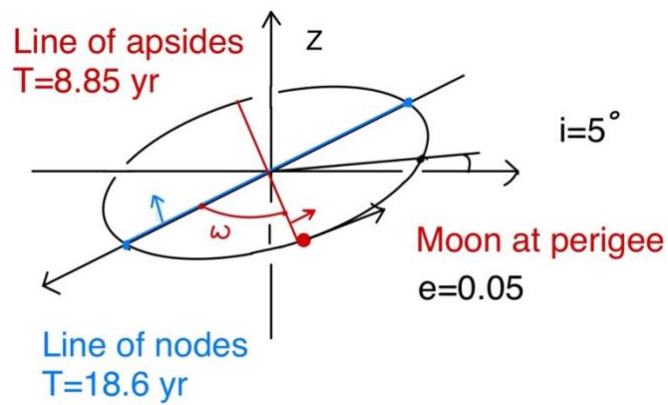

*Figure 3 Lunar precession. The Moon's orbit is an ellipse with eccentricity $e \simeq 0.05$ and inclination of $i \simeq 5°$ relative to the ecliptic plane (normal to the z axis). The ellipse undergoes two slow precessions: the orbital plane precesses around the z axis such that the line of nodes (the intersection of the orbital plane with the ecliptic) regresses westward with a period of about 18.6 years, while the ellipse precesses in its plane such that the line of apsides (joining perigee and apogee, the points nearest to and farthest from the Earth, respectively) advances eastward with a period of about 8.85 years.*

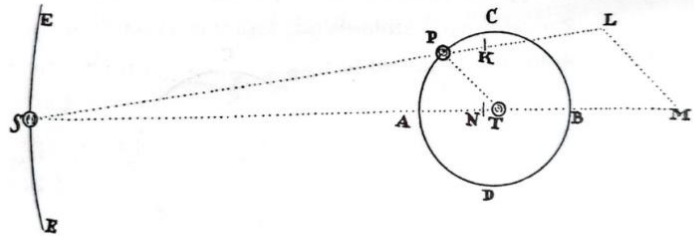

*Figure 4 A figure from the Principia (Book I, proposition 66) describing a gravitating system consisting of three bodies: S, T and P.*

Crack in inverse-square law and its return. The scientists that followed Newton continued to struggle with the lunar apsidal precession. On 15 November 1747, the French mathematician Alexis Clairaut presented a talk on this subject in Paris. He reached the conclusion that Newton's inverse-square law cannot explain this precession. Realizing that natural laws are often approximate, he experimented with modifying this law.

Euler and d'Alembert were trying to resolve the puzzle at the same time. In a reply letter to Clairaut, Euler wrote that he had already pointed out that the inverse-square law is insufficient to explain the motion of the Moon, yet the suggested modification is incompatible with the motion of Mercury.

On 17 May 1749, before this priority dispute was settled, Clairaut announced that at last, he *was able* to explain the Moon's precession with Newton's uncorrected law. By considering a higher correction in the ratio of the Moon period over Earth's, he was able to resolve most of the discrepancy.

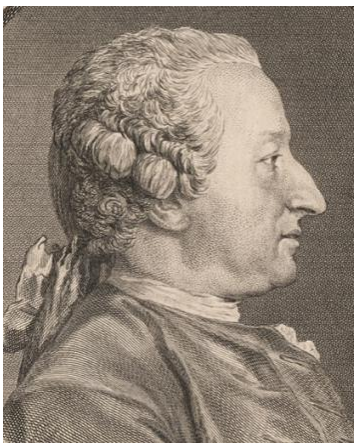 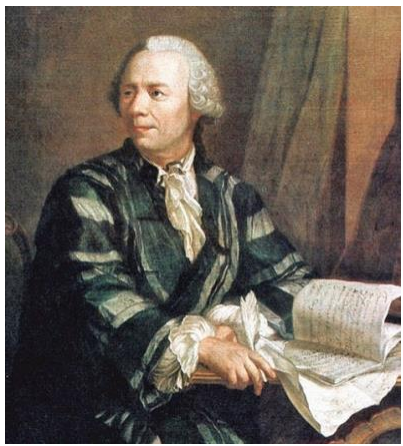 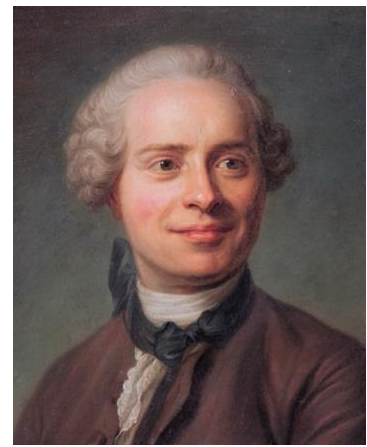

*Figure 5 Scientists who addressed the apparent discrepancy between lunar precession and the inverse-square law: Alexis Clairaut (1713–1765), Leonhard Euler (1707–1783) and Jean le Rond d'Alembert (1717–1783).*

In this way, the Moon's precessions were explained as a residual three-body effect, and the inverse-square law withstood a stringent test. This episode reaffirmed Newtonian gravitation while revealing the analytical fragility of multi-body motion. See (Bodenmann, 2010) for a detailed account.

The general three-body problem. Once the validity of the law of gravitation had been firmly established, and inspired by the success of the general solution to the two-body problem, attention turned naturally from the Earth–Moon–Sun system to the general three-body system, allowing arbitrary masses and initial conditions.

In 1760, Euler found a general solution to a related problem, that of a body moving under the influence of two fixed centers of gravitational attraction (Euler, 1760). Using elliptic coordinates, the problem was seen to be integrable.

For the fully interacting three-body case, two exact special solutions were found, both for general masses but with specific initial conditions. Euler discovered the rotating collinear solutions (Euler, 1767), while Lagrange found the rotating equilateral solutions (Lagrange, 1772). In these solutions, each one of the bodies moves on a Keplerian orbit around the center of mass, such that the entire configuration rotates—and possibly rescales—in unison. These two classes of solutions generalize the notion of Lagrange points (Figure 6). To date, they remain the only known closed-form solutions of the general three-body problem.

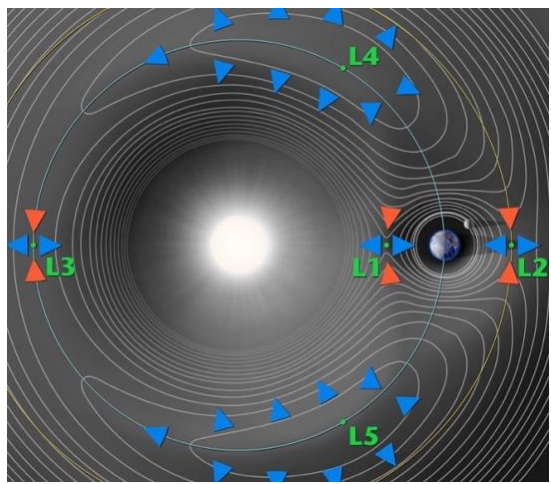

*Figure 6 The Five Lagrange points – solutions of the restricted three-body problem known in closed form. The three collinear configurations are due to (Euler, 1767), and the equilateral ones are due to (Lagrange, 1772). These solutions generalize to arbitrary mass sets (the general problem) and to eccentric orbits (image credit: NASA).*

Lagrange introduced a change of variables, replacing the standard three position vector variables by the side lengths of the triangle defined by the three bodies and by other variables. This formulation reduces the order of the equations of motion as a system of differential equations. The discovery of the equilateral solution was a by-product of this formulation.

Hierarchical limits. Just as success leads to a period of rapid expansion, eventually, expansion encounters a boundary and slows down. In celestial mechanics, the success of the gravitational law and the complete solution of the two-body problem led to rapid expansion—from the planets to the Moon, and to ambitious attempts to solve the general three-body problem. However, a general solution was not found, not even within Lagrange's reduced formulation in terms of triangle geometry, thereby signaling a boundary for expansion.

In situations when a general solution cannot be found, solutions for more special cases are sought. For the three-body problem, there are two useful special cases (Figure 7). The first is the planetary case, where a heavy mass, such as the Sun, is orbited by two lighter masses, such as two planets (a mass hierarchy). The second is the lunar case, where a close binary, such as the Earth and Moon, orbits a distant third body, such as the Sun (orbital hierarchy).

Perturbation theory makes it possible to determine corrections to the above-mentioned exact special solutions, thereby extending their range of applicability. In this way, more realistic systems—featuring hierarchies in mass or in orbital configuration—can be understood.

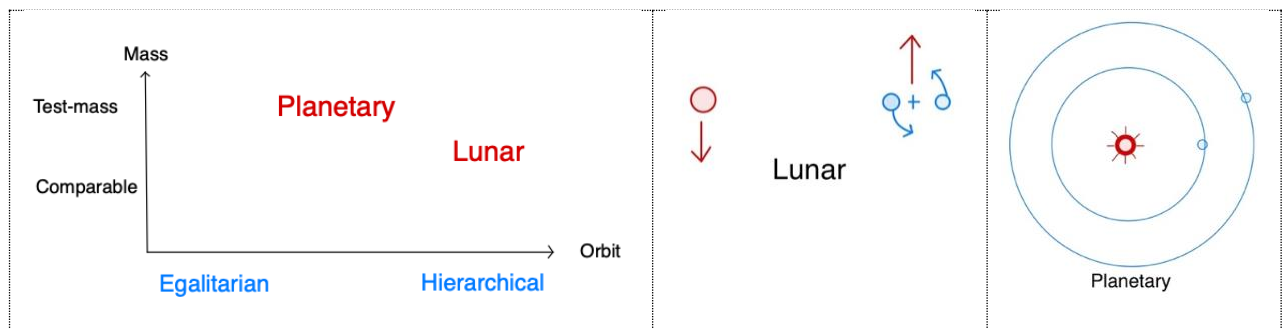

*Figure 7 Three-body regimes. Parameter space of the three-body problem, with regimes and limiting cases indicated and illustrated schematically.*

The lunar case was already discussed above. We now turn to two developments regarding the planetary limit. During the 1770s, Laplace and Lagrange studied the planetary limit and demonstrated the robustness of the planets' semi-major axis. This was interpreted as relevant for the stability of the Solar System. For more details on this, and on later developments regarding the stability of the Solar System, see appendix A.

Symplectic formulation of mechanics. In 1808, Lagrange was asked to review recent work by Poisson on the planetary limit of the three-body problem. To lowest order, each planet follows a Keplerian orbit; the objective was to determine the cumulative effect of weak mutual interactions. While 72-year-old Lagrange studied the work of 27-year-old Poisson, he recalled his own work from 1775 and 1779 on the method of variation of constants and realized that he could economize the computation of planetary perturbations. He introduced the slowly varying orbit parameters (the osculating elements), and found that he could describe their time evolution in terms of an antisymmetric product—now called the Lagrange bracket (Lagrange, 1808). He then extended this notion to general mechanical systems (Lagrange, 1809).

(Poisson, 1809) improved on this definition by introducing the Poisson bracket, the inverse of the matrix of Lagrange brackets. As explained in (Lagrange, 1810), Poisson brackets have the advantage of being defined for any pair of dynamical variables, whereas Lagrange brackets depend on a choice of a maximal set of them. For this reason, the Poisson brackets replaced Lagrange's; see the reviews (Marle, 2009), (Iglesias, 2013).

Poisson brackets underlie the symplectic formulation of mechanics: they reveal an invariant symplectic structure and lead to integral invariants (over ensembles). This became central to the foundational Hamiltonian formulation that followed and later provided the language for modern analyses of chaos and stability.

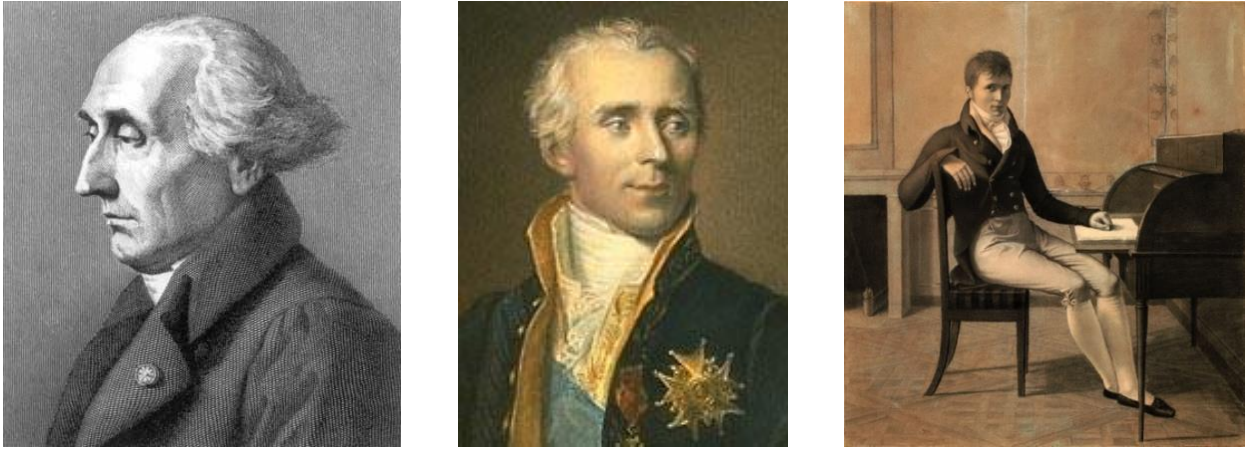

*Figure 8 From left to right: Joseph-Louis Lagrange (1736–1813), Pierre-Simon Laplace (1749–1827) and Siméon Denis Poisson (1781–1840) shown in 1804. Euler and Lagrange obtained special solutions to the three-body problem; Lagrange and Laplace established a result on the stability of the Solar System; and Poisson and Lagrange developed the symplectic formulation of mechanics.*

Elimination of nodes. A generation later, Jacobi revisited with the three-body problem. In (Jacobi, 1836), he studied a test-mass in the background of a circular massive binary—thcircular 3BP—and, in the rotating frame, introduced an energy-like conserved quantity now known as the Jacobi invariant. In (Jacobi, 1843), he further improved Lagrange's reduction by lowering the order of the equations by one via the elimination of the nodes, at least in the case of orbital hierarchy.

An important corollary to Jacobi's invariant awaited Hill's work (Hill, 1877), (Hill, 1878). Hill showed that Jacobi's invariant defines an allowed region for the motion of a test-particle (such as the Moon) and that motion confined to a bounded region is necessarily bounded—and hence stable. These allowed regions are now called Hill regions of stability.

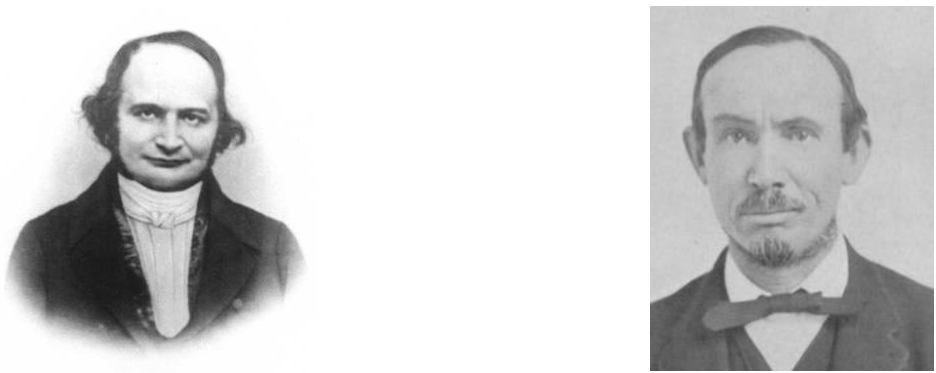

*Figure 9 Carl Gustav Jacob Jacobi (1804–1851), on the left, and George William Hill (1838–1914).*

## Chaos

Sometimes a goal cannot be reached simply because it never existed. In 1885, King Oskar II of Sweden and Norway, advised by the mathematician Mittag-Leffler, announced a mathematical competition. The first of the invited subjects was "to expand the coordinates of each particle [of the Newtonian N-body problem] in a series", see e.g. (Gray, 2013, p. 267). After two centuries of searching, this phrasing reflected the growing impression that a closed-form solution was unlikely, and hence a more modest goal was assigned.

Henri Poincaré, then 31 years old, was drawn to the challenge. Eventually, he submitted an essay on the subject, and even though it did not meet the assigned goal, it was recognized as exceptionally original, won him the prize, and made him famous. During preparation for publication, Poincaré realized that a substantial revision was needed. This revised version (Poincaré, 1890), submitted on 5 January 1890, will be described below. For more on the story of the prize and the resubmission, see (Rågstedt), (Barrow-Green, 1996).

Poincaré chose to focus on a simplified and distilled limit of the three-body problem (3BP) that preserves the key challenge. Specifically, he studied a test-particle moving under the influence of a circular binary and within the binary's plane – a case known as the planar restricted circular 3BP. Furthermore, in order to employ perturbation theory, he assumed a mass hierarchy within the binary and denoted the binary mass ratio by $\mu \ll 1$. In the small $\mu$ limit, the system has two kinds of periodic solutions, planetary and lunar-like. Poincaré studied small perturbations to these periodic solutions. He described this approach as "the only breach … to penetrate a fortress hitherto deemed unassailable" (Poincare, 1892-9, p. 82).

During this work, Poincaré developed several novel methods that are known today as Poincaré sections, Poincaré recurrences and characteristic exponents. He also developed the Poincaré–Lindstedt method that enables one to determine perturbations to a periodic trajectory, where the frequency is allowed to depend on the small parameter. Finally, he developed asymptotic (diverging) power series, which are central to modern Quantum Field Theory. Poincaré introduced them a few years earlier in a different context, and was surprised to encounter them here. In this case, the role of the small parameter is played by $\mu$. He even described non-perturbative terms, which he called "infinitesimals of infinite order" (Figure 10).

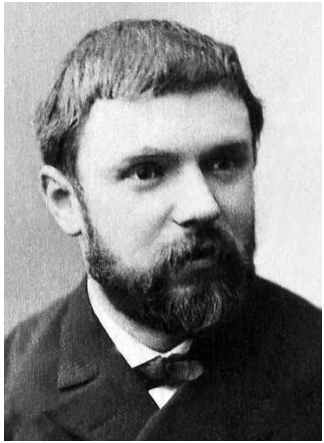 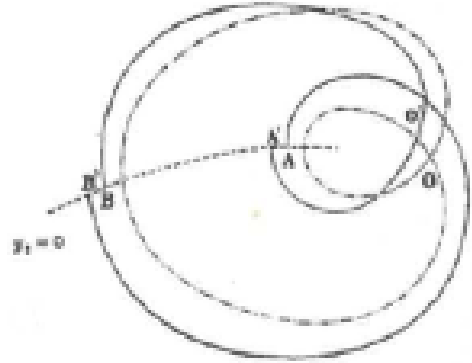

*Figure 10 Left: Henri Poincaré (1854–1912), shown around 1887. Right: a non-perturbative sign of non-integrability from Poincare's essay: AA' and BB' denote the mismatch between the stable and unstable manifolds.*

Each one of these novelties deserves a dedicated discussion, and indeed, based on them, in the ensuing decade, Poincaré wrote the three volumes of "Les méthodes nouvelles de la mécanique céleste" (Poincare, 1892-9).

Here we focus on the characteristic exponents. Poincaré considered a small deviation from the periodic orbit. The deviation satisfies a linear equation of motion, and therefore, after a full period it returns to itself up to a linear transformation. The eigenvalues of this transformation are termed the characteristic exponents. Poincaré found that some exceeded unity in absolute value, hence the deviation grows exponentially with time, at least as long as the approximation of linear deviation holds, and the solution displays an extreme sensitivity to initial conditions.

The case studied by Poincaré has only 2 degrees of freedom (planar motion), and a single conserved quantity (Jacobi's). Poincaré proved that the growing exponentials imply the striking result that the system contains no additional conserved quantities. In modern terminology, this makes the system non-integrable. This generalizes to the general three-body system, which has no conserved quantities beyond the total energy, the total linear momentum, the center of mass, and the total angular momentum.

Sensitivity to initial conditions later became the cornerstone of modern chaos theory, which developed during the 1960s-70s. Previously, it was believed that knowledge of the forces at work, together with the initial conditions would allow a deterministic prediction. This view, inspired by exact solutions to mechanical problems, such as the two-body problem, was articulated by Laplace, and became known as the clockwork universe. Exponential sensitivity to initial conditions, together with the fact that the initial conditions are never known with infinite precision, implies a limit on prediction: an increase of precision of initial conditions by a multiplicative factor (say, doubling) changes the predictive time only by an additive constant. Moreover, generic mechanical systems are believed to display such sensitivity and hence are chaotic. In this way, chaos brought about a paradigm shift in science, replacing the clockwork universe by limited predictability.

Because of limited predictability, it is believed that a general solution to chaotic systems, including the three-body problem, is not only complicated, but in fact, is impossible(!)—at least not one "resembling any known analytical tools" according to Poincaré.

Later, chaos was found to be accompanied by fractals, baroque–like geometric structures in phase-space that reveal infinite detail at ever smaller scales. In this sense, chaos can be viewed as a form of complexity, and the three-body problem as one of its simplest manifestations: the first to be studied and recognized as such. It also represents the immediate generalization of the integrable two-body problem and is therefore "simple" in this further sense.

In hindsight, how should we evaluate the journey towards a general solution to the three-body problem? Was the pursuit, misguided as it was, in vain? The author believes not. First, starting with Newton and during the first half-century thereafter, the goal was to understand the motion of the Moon, which is certainly amenable to analysis and hence a valid goal. Next, Euler, Lagrange and later Jacobi engaged with the general problem. While they did not, and indeed, could not have found the general solution, they were able to make important discoveries: they found new special exact solutions, and they dynamically reduced the problem into that part which, in hindsight, is its essential chaotic core.  Finally, two centuries after the problem's inception, Poincaré caught the first glimpse of chaos. While it is conceivable that chaos could have been discovered without attempting the misguided general solution, in reality, it was the failed search for the general solution that finally gave birth to the revolution that is chaos, and to progress. Thus, the search was conducted with sufficient good judgement that even though the goal did not exist, the search was still most fruitful.

It is fascinating to note that while Poincaré was working on Méthodes Nouvelles, he invented homology and homotopy in (Poincaré, 1895), thereby initiating algebraic topology and manifold topology, which are the core of topology, and would become a central topic for 20$^{th}$ century mathematics. This is arguably Poincaré's greatest single achievement. In the introduction to this topological work, Poincaré cites celestial mechanics among his motivations.

After Poincaré. At the beginning of the 20$^{th}$ century, physicists' attention turned to relativity and quantum mechanics, yet, important research on the three-body problem continued.

We mention a limited selection of works. (Levi-Civita, 1915) presented a Hamiltonian reduction of the general three-body problem. The quantum mechanical version of the proved essential in establishing quantum mechanics, particularly in understanding the spectrum of the Helium atom (2 electrons + nucleus = 3 bodies). In this context, (Hylleraas, 1929) presented a theory together with numerical results, and demonstrated a good fit with experimental data.

Returning to the classical domain, (Lemaître, 1952) introduced a set of symmetric coordinates, based on the side lengths of the triangle formed by the three bodies, which remain smooth even in the collinear limit. (Fock, 1954) presented a 4d higher dimensional perspective of this space. Later, (Moeckel & Montgomery, 2013) introduced a modern mathematical perspective and coined the term *shape sphere* for this space of triangle shapes, consistent with Lemaître's construction.

The onset of chaos is described by nearly integrable systems within the Kolmogorov-Arnold-Moser (KAM) theory (Kolmogorov, 1954) (Moser, 1962) (Arnold, 1963). For integrable systems with generic incommensurate frequencies, the closure of the trajectory in phase space forms an invariant torus whose dimension equals the number of degrees of freedom. The KAM theorem states that when integrability is broken by a small perturbation, most of these tori persist, changing dimension only along specific loci in phase space—a phenomenon known as the *preservation of invariant tori*. The time evolution of instability in such nearly integrable systems is described by the Nekhoroshev estimates, while the resulting drift in phase space is known as Arnold's diffusion.

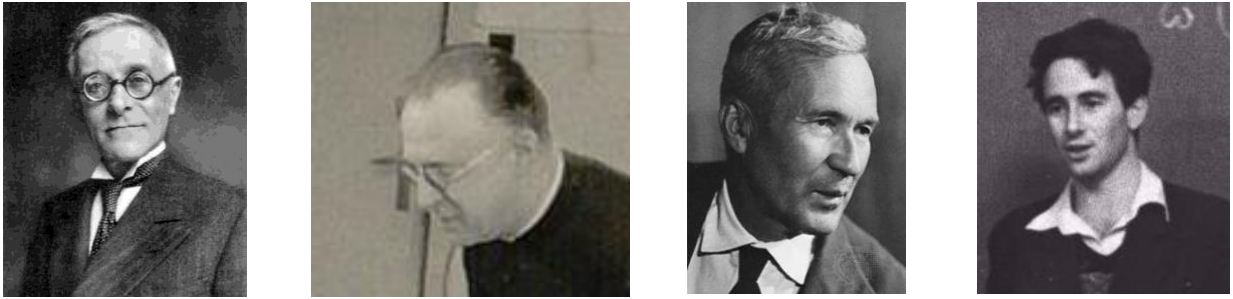

*Figure 11 Twentieth-century scientists who made significant contributions to research on the three-body problem. From left to right: Tullio Levi-Civita (1874–1941), Georges Lemaître (1894–1966), Andrey Kolmogorov (1903–1987) and Vladimir Arnold (1937–2010). Image credits: Lemaître and Arnold – Wikipedia; Kolmogorov – Encyclopaedia Brittanica.*

The book (Celletti, 2009) discusses stability and chaos in celestial mechanics, and the review (Naoz, 2016) examines the implications of interactions within hierarchical systems.

These developments show that, even as physics expanded, the three-body problem remained a central puzzle for ideas about motion, stability, and chaos.

While three-body research was long motivated by subsystems within the Solar System, in recent decades several new applications have emerged. Star clusters, with their high stellar densities, frequently host binary systems approached by a third star, leading to triple interactions. Three-body dynamics is therefore essential for understanding stellar populations in clusters. The discovery of exoplanets over the last three decades revealed numerous planetary systems with significant eccentricities and inclinations. These are explained by triple interactions involving combinations of stars and planets. Tight binary systems are now understood to be progenitors of gravitational waves, routinely detected since 2015. Their formation is believed to require a specific mechanism, and triple interaction are among the leading proposals (McMillan & Portegies Zwart, 2000). Moreover, triple interactions are implicated in producing Type Ia supernovae (Maoz, Mannucci, & Nelemans, 2014). Finally, the three-body problem was generalized beyond Newtonian gravity to other interactions: for example, the electric three-body system is relevant to Helium-like atoms; the nuclear three-body system describes light nuclei composed of three nucleons; and the harmonic three-body system, where the bodies interact through springs, has been studies in the context of dynamical systems.

## Statistical theory

Having seen how determinism gives way to chaos, we now turn to how statistical reasoning restores predictability at a different level.

The development of electronic computers enabled approximate solutions to differential equations through computer simulations, rather than analytic methods. The pioneering study (Agekyan & Anosova, 1967) simulated the non-hierarchical (or egalitarian) three-body system and was soon followed by the work of Aarseth, Valtonen, Heggie, Hut, McMillan, and others. These simulations demonstrated convincingly how random and chaotic the system's motion is (Figure 12). The conserved quantities allow a three-body system to disintegrate into a binary and a single body. Because of its ergodic nature, one may expect that anything possible will eventually occur—and indeed, simulations show that almost all systems end in disintegration of this kind.

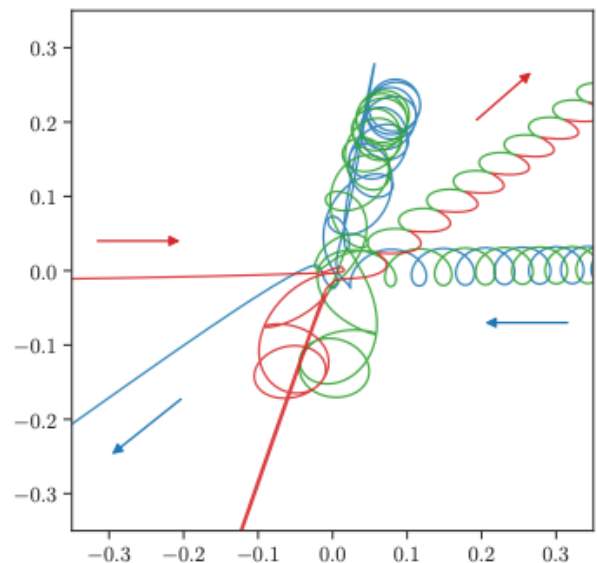

*Figure 12 The trace of a three-body time-evolution demonstrating its chaotic nature. Credit: Alessandro Trani.*

As the random-like behavior rendered deterministic prediction impossible, researchers began to collect outcome statistics. Given the masses and the conserved quantities, what are the probabilities for each of the masses to escape? What are the distributions of the escape velocity, binary parameters and decay times? Data accumulated and begged the question: could one formulate a theory that explains and predicts these statistics? It became clear that the unattainable deterministic solution was a wrong goal and it should be replaced by a statistical one.

A general framework for such reasoning—statistical mechanics—has existed since (Gibbs, 1902). It models uncertainty in the system's state by an ensemble of systems occupying different regions of phase space. Note that since the system has a small number of degrees of freedom, we are not in the thermodynamic limit (which corresponds to a large number of degrees of freedom). Accordingly, the statistical mechanics of the three-body system is somewhat different from that of a thermodynamic system. Yet, Statistical Mechanics could not be applied straightforwardly to the three-body problem, because it features the following combination of three elements: instability to disintegration, unboundedness, and a phase-space that contains both chaotic and regular regions.

(Monaghan, 1976) introduced the first statistical theory for the egalitarian three-body problem. It addressed the above-mentioned elements by introducing the "strong interaction region" R, which serves both as a cutoff (confining the system to a box) and to separate regular from chaotic motion. It renders phase space finite, and thereby phase space volume can be equated with probability in the usual way, without producing infinite probabilities. Until recently, all statistical theories were based on phase-volume in this way and contained $R$.

However, R is clearly a spurious parameter—absent from the underlying dynamics and adjustable only to fit results. As such, a more fundamental theory should be possible, one that avoids it altogether.

Flux-based statistical theory. Building on earlier work, (Stone & Leigh, 2019) advanced the statistical theory by obtaining a closed form expression for the outcome distribution, enabled by a judicious choice of integration variables, see also (Ginat & Perets, 2021).

This elegant result attracted the author's attention. However, it soon became clear that the approach was incomplete because it continued to rely on the spurious parameter R.

An ensuing examination of the foundations of the statistical theory, presented in (Kol, 2021), brought several key changes. First, it was found that R could be eliminated while keeping the outcome probabilities finite. This was achieved by introducing a diverging normalization factor that renders the probabilities well defined without any adjustable cutoff.

Secondly and most importantly, it was realized that since the system almost always disintegrates and we are interested in the outcome distribution, the relevant quantity is the decay rate distribution rather than the probability distribution. Accordingly, the statistical theory must evaluate the flux of phase-space volume, not the volume itself. The situation is analogous to a leaky container, where the exit probability per unit time of a given molecule (its decay rate) is proportional to the flux of fluid through the opening. Replacing phase-space volume by flux automatically yields finite decay rates and removes dependence on where the flux is measured (because the flux is conserved). This shift established a new statistical framework—the flux-based statistical theory.

Thirdly, the division of phase space into regular and chaotic regions was accounted for. Not every triple encounter leads to a chaotic motion; some lead to regular motion such as a flybys or a direct exchange of the tertiary with one of the binary components. The probability for chaotic absorptivity can therefore range between zero and one (we use the term probability even though the system is deterministic, since we imagine averaging over the initial binary phase). By time reversal symmetry, not every disintegration originates from chaotic motion, and accordingly, one can define a *chaotic emissivity function*. In this way, the decay rate distribution arising from chaotic motion can be written as a product of the flux distribution and the chaotic emissivity function.

The distribution of flux was found in closed form, reducing the statistical solution to determining the chaotic emissivity function. Two complementary approaches are possible. The first is empirical: measure the absorptivity function in simulations and invoke the equality of absorptivity and emissivity implied by time reversal symmetry.

Such simulations are computationally efficient because they need only track the approach of the tertiary to the binary and the ensuing few close encounters—there is no need to continue until final disintegration. The equality of chaotic emissivity and absorptivity is analogous to **Kirchhoff's law** of thermal radiation, which equates a body's absorptivity and emissivity for electromagnetic radiation.

A second approach, for future work, is phenomenologic–analytic: obtain analytic approximations for the chaotic emissivity function itself. Together with the closed-form flux distribution, any such approximation would immediately yield an analytic prediction for the outcome distribution.

Fourthly and lastly, the theory was extended to account for episodes of regular motion, where the system temporarily separates into a binary and a single star whose relative velocity is below the escape velocity. Such "sub-escape excursions" necessarily end in renewed triple encounters. The theory incorporates these by tracking the probability flow among chaotic regions, excursions, and eventual disintegrations, formulating a time-evolution equation in which the decay-rate distribution plays a central role.

The flux-based theory differs from all previous statistical theories both in foundation and in prediction, which has been tested extensively through numerical simulations. Because the outcome distribution reduces to the emissivity function, predictions require some knowledge of that function. The first major validation (Manwadkar, Trani, & Leigh, 2020) (Manwadkar, Kol, Trani, & Leigh, 2021) applied an emissivity-blind assumption to predict the escape probabilities (the probability for each one of the three masses to escape) and achieved agreement at the 1% level, compared with only $\sim 7 - 10\%$ for earlier volume-based theories (Figure 13).

A second, more stringent validation came from numerical measurements of a bivariate distribution of the chaotic emissivity function. This quantity was then used, via the flux-based theory, to predict a detailed bi-variate outcome distribution, which was compared with a direct measurement of the outcome distribution (Manwadkar, Trani, & Kol, 2024) and showed excellent agreement (Figure 14). Altogether, numerical tests demonstrate a substantial leap in predictive accuracy, establishing the flux-based theory as the most precise statistical theory of the three-body problem to date.

| Masses($M_\odot$) | Ejected | $N_s$ | $P_s$ | K20 | SL19 | VK06 |
|---|---|---|---|---|---|---|
| 15, 15, 15 | 15 (0) | 45208 | 0.328 | 0.333 | 0.333 | 0.333 |
|  | 15 (1) | 46655 | 0.339 | 0.333 | 0.333 | 0.333 |
|  | 15 (2) | 45771 | 0.333 | 0.333 | 0.333 | 0.333 |
| 12.5, 15, 17.5 | 12.5 | 89428 | 0.572 | 0.562 | 0.639 | 0.492 |
|  | 15 | 42525 | 0.272 | 0.279 | 0.247 | 0.305 |
|  | 17.5 | 24429 | 0.156 | 0.159 | 0.114 | 0.203 |
| 12, 15, 18 | 12 | 98807 | 0.618 | 0.610 | 0.696 | 0.525 |
|  | 15 | 40136 | 0.251 | 0.257 | 0.217 | 0.293 |
|  | 18 | 20881 | 0.131 | 0.133 | 0.087 | 0.182 |
| 10, 10, 20 | 10 | 67278 | 0.4736 | 0.4805 | 0.492 | 0.463 |
|  | 10 | 69555 | 0.4896 | 0.4805 | 0.492 | 0.463 |
|  | 20 | 5233 | 0.0368 | 0.0390 | 0.015 | 0.074 |
| 10, 15, 20 | 10 | 113627 | 0.784 | 0.784 | 0.872 | 0.649 |
|  | 15 | 22988 | 0.159 | 0.159 | 0.103 | 0.236 |
|  | 20 | 8306 | 0.057 | 0.057 | 0.025 | 0.115 |
| 10, 20, 20 | 10 | 108978 | 0.875 | 0.880 | 0.948 | 0.726 |
|  | 20 | 8055 | 0.065 | 0.060 | 0.026 | 0.137 |
|  | 20 | 8055 | 0.065 | 0.060 | 0.026 | 0.137 |
| 8, 21, 21 | 8 | 86959 | 0.943 | 0.956 | 0.987 | 0.8158 |
|  | 21 | 2813 | 0.031 | 0.022 | 0.006 | 0.0921 |
|  | 21 | 2813 | 0.031 | 0.022 | 0.006 | 0.0921 |
| 5, 15, 25 | 5 | 41960 | 0.9712 | 0.9872 | 0.9969 | 0.8796 |
|  | 15 | 1108 | 0.0256 | 0.0108 | 0.0028 | 0.0895 |
|  | 25 | 137 | 0.0032 | 0.0020 | 0.0003 | 0.0309 |

*Figure 13 Measured escape probabilities $P_s$ for different mass sets, compared with theoretical predictions. Probabilities are computed after selecting for ergodic escapes, and $N_s$ denotes the absolute number of selected time evolutions, out of a total of one million simulations per mass set. Three theoretical predictions are shown: K20 represents the flux-based theory, while SL19 and VK06 are volume-based, as described in (Manwadkar, Kol, Trani, & Leigh, 2021). Adapted from the same source.*

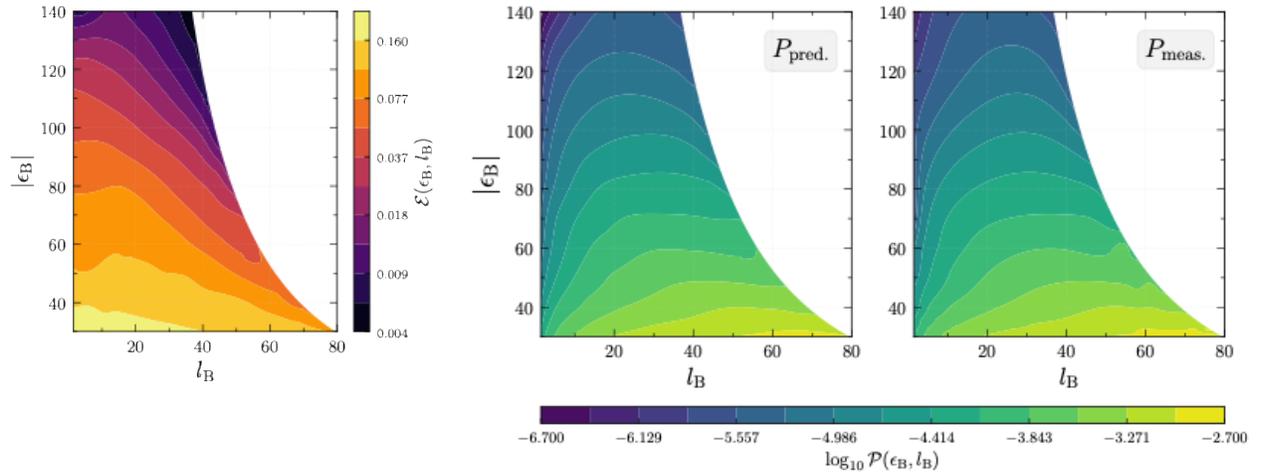

*Figure 14 Successful prediction of detailed bivariate outcome distribution by the flux-based theory. Left: measured emissivity function. Right: predicted and measured outcome distribution of binary parameters (Manwadkar, Trani, & Kol, 2024).*

Dynamical reduction. The flux-based statistical theory of non-hierarchical triples has also led to progress on a different aspect of the three-body problem, namely, its natural dynamical reduction (Kol, 2023). By this we mean a change of dynamical variables that leads to an improved formulation of the system, rather than its solution. Any three-body configuration defines both a triangle, and its orientation in space. Accordingly, the dynamical variables can be decomposed into triangle geometry variables (shape and size) and orientation variables. The geometry variables describe the motion of an abstract point in a curved 3d space, subject to a potential-derived force and a magnetic-like force with an effective monopole charge (Figure 15).

The dynamics of the orientation variables, on the other hand, resemble the familiar Euler equations for a rotating body. In this way, the three-body problem was shown to be naturally equivalent to a point moving in an abstract 3d geometry space coupled to a non-rigid rotor (a rotating body with a time-dependent inertia tensor).

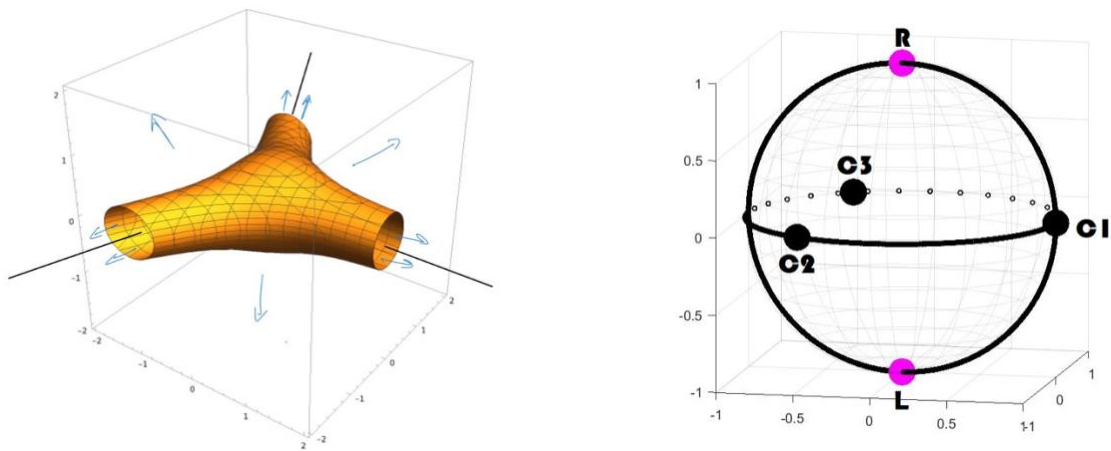

*Figure 15 Triangle geometry variables – the basis of a new formulation. Left: triangle-geometry space and a represerntative equipotential surface shaped like a pipe joint. The entire planar system is described by a single point moving within this surface, where each point corresponds to the size and shape of the instantaneous triangle formed by the three bodies. Right: Shape sphere. The angular coordinates of the triangle-geometry space describe the triangle's shape rather than its size. The poles correspond to right- and left-handed triangles, while the equator represents collinear configurations, including three coincident ones. Adapted from (Kol, 2023).*

## Conclusion

We have recounted the story of the three-body problem from Newton to the present — its central role in the development of science, and its transformation from a quest for exact trajectories to a study of collective patterns and statistical laws. We did not present all the details, but rather, strove to provide a wide perspective, together with references for further reading.

We saw how a study of this system by Poincaré heralded a paradigm shift: the clockwork universe and determinism were replaced by chaos and limits on prediction. Along the way, it produced several great puzzles, which drove science, and developed into complete scientific dramas. The first phase of the three-body story started with Newton and focused on the search for a deterministic prediction. When describing this phase of the story, we mentioned: lunar precession and physical perturbation theory by Newton; doubts over the inverse-square law and their clarification during 1747–1749 by Clairaut, Euler and d'Alembert; the robustness of a planetary semi-major axes (and the stability it suggests) during the 1770s by Laplace and Lagrange; and, finally, the discovery of the symplectic formulation of mechanics during 1808–1810 by Lagrange and Poisson.

The second phase of the story started with Poincaré and the first signs of chaos. During it, the three-body problem motivated Poincaré in creating topology; it provided an important early confirmation of quantum mechanics when applied to the Helium atom; and finally, it was a prime example for the development of the theory of nearly integrable systems (KAM).

The review concluded with the recent formulation of a flux-based statistical theory for the egalitarian three-body system. It changed the foundation of earlier three-body statistical theories, and showed a leap in agreement with computer simulations. The flux-based theory reduces its statistical solution to the evaluation of the chaotic absorptivity function, and in this sense, cracks the problem.

Overall, it was possible to appreciate how rich and complex the three-body problem is: it has numerous aspects that inspired scientific stories, it contains both perturbative and ergodic behaviors, and finally the chaotic behavior is as complex as one could imagine. And all of this waits just outside the door of the familiar and well-tamed two-body problem.

Open problems. The field is rich with open problems and goals and we mention some: advancing the flux-based theory through analytical approximations of the emissivity function; developing a statistical theory for hierarchical triples thereby extending statistical methods beyond the egalitarian regime; and lastly addressing Smale's

22Problem No. 6 for 21$^{st}$ century mathematics that involves rigidly rotating solutions (central configurations) of Newtonian many-body systems.

Altogether, more than three centuries since its inception, the three-body system continues to inspire research across diverse scientific fields. These range from the astrophysics of star clusters, exoplanets and tight binaries, to computer simulations, from physics to mathematics, and from celestial mechanics to dynamical systems. The three-body problem, indeed, remains vital.

## Acknowledgements

The author thanks P. Iglesias-Zemmour and A. Trani for helpful discussions. He acknowledges support by the Israel Science Foundation (grants nos. 1345/21 and 2058/25) and by a grant from Israel's Council of Higher Education.This essay was originally invited by the journal *Inference* for a general scientific readership. The journal was discontinued in August 2022, when the essay was already in an advanced stage of preparation.

## Appendix A: Stability of the Solar System

Before Newton, the heavens were considered unchanging and eternal, and the planets were thought to move along invisible tracks (or spheres), see e.g. (Westfall, 1983, p. 6). Once Newton realized that each planet was held in orbit solely by the Sun's gravity, the possibility arose that the orbit of the Earth, or any other planet, might change as a result of the gravitational influence of a third body. Over long timescales, such interactions could lead to the ejection of a planet from the Solar System ("runaway") or to a collision of two planets.

This must have been a disturbing thought. Newton wrote: "the Planets move one and the same way in Orbs concentrick, some inconsiderable Irregularities excepted, which may have arisen from the mutual Actions of Comets and Planets upon one another, and which will be apt to increase, till this System wants a Reformation" (Newton, 1704). In fact, according to (Hoskin, 2001), in Newton's worldview "God demonstrated his continuing concern for his clockwork universe by entering into what we might describe as a permanent servicing contract".

When assessing possible changes to planet orbits, one must consider the weak but cumulative interplanetary forces. Because these forces are small, it suffices to study two planets (and the Sun) at a time —the planetary limit of the three-body problem. For reviews of the stability of the Solar System in this context, see (Laskar, 2013) and (Tremaine, 2011).

The weakness of interplanetary forces also suggests focusing on their long-term effects by averaging over complete orbits —a procedure known as the secular approximation. Newton had already introduced a version of this method for the Moon by considering its mass to be distributed along its orbit (Gutzwiller, 1998), (Chandrasekhar, 1995). This approach is also known as Gauss's averaging method, see e.g. (Murray & Dermott, 2000, p. 293).

Beyond interplanetary forces, a planet's orbit may be disturbed by the passage of a third body—whether a comet, an asteroid, or another improbable object. The disturbance may arise through gravitational attraction, direct collision, or even an explosion. Such encounters are inherently unpredictable.

A century after Newton, during the 1770's, in yet another instance of scholarly competition that benefits science, Laplace and Lagrange proved that within the secular approximation the semi-major axis of a planet remains constant, see (Laplace, 1773; Iglesias, 2013) (Lagrange, 1781), as well as (Laskar, 2013) and references therein. In hindsight the reason is clear: the secular approximation replaces the perturbing body with a time-independent, orbit-averaged mass distribution. By a standard theorem of mechanics, time independence implies conservation of energy for the perturbed planet. Because the orbital energy determines the semi-major axis, that axis must remain constant. This result was sometimes described as guaranteeing the stability of the Solar System. Indeed, it forbids certain types of instability, such as the ejection of a planet, although it relies on the validity of the secular approximation and cannot exclude instabilities beyond its assumptions.

Over time, stronger notions of stability were studied. Within the circular-binary test-mass approximation, the Hill region (described in the "Elimination of nodes" section) protects from runaway without assuming the secular approximation. (Poincaré, 1890) showed that a non-periodic system evolving in a bounded phase space must eventually return arbitrarily close to its initial state—a form of stability known as Poincaré recurrence. Finally, KAM theory guarantees a degree of stability in the nearly integrable limit.

Ultimately, none of these considerations ensures absolute stability of the Solar System. Computerized simulations of the Solar System—modeling the Sun and planets as point masses under Newtonian gravity—show that over tens of millions of years the planetary positions, especially that of Mercury, become unpredictable due to exponential sensitivity to initial conditions (Laskar, 2013; Tremaine, 2011). Moreover, models of Solar System formation suggest that earlier planetary siblings of Earth may once have existed and were later lost, revealing another kind of instability.

This means that the Solar System, just like all natural systems, is not perfectly stable, and one is required to come to terms with uncertainty.